\documentclass[journal=apchd5,manuscript=article]{achemso}
\usepackage{hyperref}

\usepackage{multicol}
\usepackage[T1]{fontenc}
\usepackage{amsmath}
\usepackage{ulem}

\usepackage{graphicx}
\usepackage{color}
\definecolor{OliveGreen}{rgb}{0,0.6,0}
\usepackage{dcolumn}
\usepackage{bm}
\usepackage{amssymb,float}
\usepackage{cancel}
\usepackage{ulem}
\usepackage{physics}
\newcommand{\Z}{\mathbb{Z}}

\author{R. Banerjee}

\author{T.C.H. Liew}
\email{timothyliew@ntu.edu.sg}
\affiliation{Division of Physics and Applied Physics, School of Physical and Mathematical Sciences, Nanyang Technological University, Singapore 637371, Singapore}

\title[Artificial Life in an Exciton-Polariton Lattice]
  {Artificial Life in an Exciton-Polariton Lattice}


\begin{document}

\begin{abstract}
We show theoretically that a lattice of exciton-polaritons can behave as a life-like cellular automaton when simultaneously excited by a continuous wave coherent field and a time-periodic sequence of non-resonant pulses. This provides a mechanism of realizing a range of highly sought spatiotemporal structures {\it under the same conditions}, including: discrete, oscillating, and rotating solitons; breathers; soliton trains; guns; and chaotic behaviour. These structures can survive in the system indefinitely, despite the presence of dissipation and disorder, and allow universal computation.
\end{abstract}

\section*{Introduction}
\label{section:Introduction}
 Exciton-polaritons are quasiparticles typically formed in microcavities where light hybridizes with excitons contained in quantum wells. As a result of this hybridization, (exciton)-polaritons have been used to study a variety of fundamental nonlinear effects, with a general motivation of reaching optoelectronic devices~\cite{Sanvitto2016}. To give a few examples, a large body of work was focused on the creation and control of polariton solitons~\cite{Amo2011,Grosso2012,Sich2012,Sich2014,Goblot2016,Walker2017,FlayacSolnyshkov2011}, which were conjectured to play a role in devices~\cite{Pinsker2014,Chana2015,Walker2015} with some soliton logic gates constructed~\cite{Cancellieri2015}. In parallel, polaritons were found to form spatial patterns~\cite{Christmann2012,Cristofolini2013,Saito2013,Whittaker2017,DiazCamacho2018,KeelingBerloff2008}, where topologically stable structures were considered as a memory~\cite{Ma2017a,Ma2017b} and pattern transitions could effectively compose switches~\cite{Lewandowki2017}. There has also been a growing interest in studying polaritons in periodic potentials~\cite{Amo2016}, which have allowed the routing of polaritons~\cite{Flayac2013,Marsault2015}, gap and lattice solitons~\cite{Tanese2013,CerdaMendez2013,Ostrovskaya2013}, and the formation of spin-ordered patterns~\cite{Ohadi2015} for information processing.

While the aforementioned works are promising for polaritonic devices in principle, a complication in applying the aforementioned phenomena is that they have been found under different conditions in different parameter ranges. Here, we consider whether solitons, stable structures, and evolving patterns can co-exist {\it under the same conditions}, together with the analogues of a variety of other soliton-related structures studied separately in the literature, including: oscillating solitons and patterns~\cite{Manni2011,Christmann2014,Gavrilov2018}; guns~\cite{Flayac2012} and soliton trains~\cite{Pinsker2014,Goblot2016,Chana2015}; soliton explosions~\cite{Xue2014}; and backward radiation emitting solitons~\cite{Skryabin2017}. We also aim for polariton solitons and related structures to exist indefinitely~\cite{Ostrovskaya2012}~\cite{Ma2017a}, beyond the finite polariton lifetime, and survive the presence of noise and disorder. To reach these aims, we attempt to associate a polariton lattice to a cellular automaton.

Cellular automata emerged as one of the first definitions of artificial life~\cite{Langton2986}, where they showed how remarkably complex behaviour associated to living organisms such as movement, growth, and replication can appear from apparently simple update rules applied on a lattice. The most commonly studied versions operate with square lattices, with each lattice site existing in one of two states typically referred to as ``alive'' or ``dead''. An update rule is applied repeatedly, where the state of each lattice site is updated depending on its own state and the state of its neighbours. The update rule defines the complexity of the corresponding automaton, where four classes of increasing complexity exist ~\cite{Stephen1983,Von1966,Wolfram2002,Wolfram1986,Wolfram1991,Wolfram1984}. Sufficiently complex automata are known for forming spatial patterns, self-localized structures (i.e., solitons), gliders or spaceships (i.e., propagating solitons), breathers (i.e., oscillating solitons) and guns (that generate soliton trains). Cellular automata have applications in image processing~\cite{ Davis1975,Chowdhury1994,Chang2004} and those belonging to the most complex class (class 4) are typically universally (Turing) complete ~ \cite{Completeness_CAs}. Life-like automata are defined as those where the update rule is based on the number of neighbouring alive states, independent of their relative position. Famous examples include Conway's life and ``Life without death'', which are universally complete class 4 automatons.

There are reasons to expect that polaritons could operate as life-like automata. First, they can be confined in square lattices~\cite{CerdaMendez2013,Kim2011,Winkler2015}. Second, under continuous near-resonant coherent excitation, polaritons exhibit bistability~\cite{Baas2004,Goblot2016}, such that each site in a lattice would be in a high intensity (i.e., ``alive'') or low intensity (i.e., ``dead'') state for as long as the coherent excitation is maintained~\cite{Cerna2013}. Furthermore the nonlinearity of polaritons suggests a potential for non-trivial behaviour, as the aforementioned works have demonstrated, however, it is a highly non-trivial question as to whether the nonlinearity can result in any complex automaton rules.

In addition to continuous near-resonant excitation, we consider the effect of non-resonant pulses applied to the system. Each pulse is found to initiate one update according to a life-like automaton rule, making use also of the spin-dependent interactions between polaritons~\cite{Takemura2014} and recently realized spin-orbit coupling in lattices~\cite{Sala2015,Whittaker2018}. This provides a platform for merging many of the separately studied soliton-related polariton phenomena in microcavities and such combination shows most clearly their prospects for information processing: we find that {\it polariton solitons are universally complete}.

\section*{Scheme}
\label{section:Scheme}
Figure ~\ref{fig:Fig1}(a) shows a schematic illustration of a square lattice of polariton resonators (e.g., micropillars \cite{Amo2016}). Each resonator represents a ``cell'' supporting two polariton spin components ($\sigma_\pm$), represented by the wavefunction $\psi_{n\pm}$, which evolves according to \cite{GippiusShelykh2007}:
\begin{align}
i\frac{\partial\psi_{n\pm}}{\partial t}=\left(-\Delta+|\psi_{n\pm}|^2+\alpha_2|\psi_{n\mp}|^2+iP_\pm(t)-\frac{i}{2}\right)\psi_{n\pm}+J\sum_{\langle m\rangle}\psi_{m,\mp}+F_\pm
\label{eq:GP}
\end{align}

$\Delta$ represents an energy detuning between a driving laser field, with circularly polarized components $F_\pm$, and the resonant polariton energy. Spin-dependent interactions are accounted for, where the interaction strength between polaritons of parallel spin is scaled to unity (through appropriate choice of scale of $\psi_{n\pm}$) and $\alpha_2$ is the relative strength of interactions between polaritons with antiparallel spin. $P_\pm(t)$ represents a gain in the system \cite{KeelingBerloff2008}, i.e., a non-resonant pulse with different spin components. The term $-i/2$ represents polariton dissipation (our time unit is the inverse dissipation rate).

We assume that each cell is coupled to its eight nearest neighbours  (see Fig.~\ref{fig:Fig1}(b)) through a spin-orbit coupling. In principle the spin-orbit coupling can be tuned through the design of the microcavity structure in the intermediate region between cells~\cite{Sala2015,Whittaker2018}. It has also been shown that polaritons propagating in channels rotate their spins as they propagate \cite{AntonMorina2015}, allowing a design of a full inversion scheme of coupling which we assume between lattice sites here.  We also assume an equivalent coupling for all neighbours (which requires that the effective potential for diagonal connections is engineered differently for horizontal/vertical connections). Although we consider here a tight-binding representation of polaritons, as is common for micropillar arrays ~\cite{Sala2015}, our results can also be reproduced with a continuous description (see sec.\ref{section:Continuous model}).

It is instructive to consider first the single cell behaviour ($J=0$). In the stationary limit (with no pulse; $P_\pm(t)$=0), the stationary states of Eq.~(\ref{eq:GP}) excited by a $\sigma_+$ circularly polarized field ($F_-=0$) show a well-known S-shaped dependence~\cite{Baas2004}(see Fig.~\ref{fig:Fig1}(c)) of the polariton intensity $|\psi_+|^2$ on the driving intensity $|F_+|^2$, given by $\left[\left(-\Delta+n_{+}+\alpha_2 n_{-}\right)^{2}+1/4 \right]n_{+}=|F_+|^{2}$, where $n_\pm=|\psi_\pm|^2$. We fix $F_+$ throughout so as to remain in the bistable regime, where $\psi_+$ in each cell must adopt either a high intensity (``alive'') or low intensity (``dead'') state in the stationary limit. The S-shaped curve depends on the detuning, $\Delta$, which is effectively renormalized by the implantation of a population of $\sigma_-$ polarized polaritons via the $\alpha_2$ dependent term in Eq.~(\ref{eq:GP}) (see Fig.~\ref{fig:Fig1}(c)).

In the limit of finite but small coupling between cells ($J\ll1$), the coupling can be considered as a perturbation. It enters as an effective driving for $\sigma_-$ polarized polaritons, where the field $\psi_{n-}$ is driven by $J\sum_{\langle m\rangle}\psi_{m,+}$, according to Eq.~(\ref{eq:GP}). It is important that this field depends on the state of neighbouring cells, such that each cell is influenced based on how many of its neighbours are dead or alive (see Fig.~\ref{fig:Fig1}(b)). In the perturbative limit, stationary states of $\sigma_-$ polarized polaritons are given by $\psi_{m-}=F_\mathrm{eff}/\left(\Delta-|\psi_{n-}|^2-\alpha_2|\psi_{n+}|^2+\frac{i}{2}\right)$,
where $F_\mathrm{eff}=J\sum_{\langle m\rangle}\psi_{m+}+F_-$. We focus first on realizing the Life without death cellular automaton, which is characterized by the behaviour that a dead cell becomes alive if and only if it has three alive neighbours. All alive cells remain alive, such that any pattern developing in this automaton remains fixed. Figure ~\ref{fig:Fig1}(d) (lower points) shows the stationary states of $\psi_{n-}$ for the case where a cell is initially dead and has different numbers of alive neighbours under the condition $P_{\pm}(t)=0$  and $ F_-=0$. Choosing now a finite $F_-$ will shift the stationary states. We choose $F_-=-J\left(3\psi_{\rm{alive}}+5\psi_{\rm{dead}}\right)$, where $\psi_{\rm{alive}}$ and $\psi_{\rm{dead}}$ represent the expected wavefunction of alive and dead neighbouring cells. This field, which will be fixed throughout, shifts the intensity of $\sigma_-$ polarized polaritons to zero when a given cell has exactly three neighbours alive and otherwise leaves a finite intensity (Fig.~\ref{fig:Fig1}(d)). Recalling Fig.~\ref{fig:Fig1}(c), the S-shaped curve describing stationary states of $\sigma_+$ polarized polaritons should now be shifted if a cell has any number of neighbours alive other than three.

Having modified the stationary states of a cell in a neighbour-dependent way, we now consider the action of a non-resonant pulse, $P_\pm(t)=P_\pm e^{-\Gamma_R t}$. Physically, the pulse excites an exciton reservoir and we take the pulse as a decaying exponential to represent decay of the reservoir (with decay rate $\Gamma_R$). The two spin components of the pulse serve different purposes. The $\sigma_-$ component amplifies the up until now weak population of $\sigma_-$ polaritons, so that their neighbour dependent intensity has a more significant effect. At the same time, the $\sigma_+$ component attempts to switch the state of $\sigma_+$ polarized polaritons to the alive state. Remarkably, we find that for well-chosen parameters (see~sec. \ref{section:Other cellular automata}) such switching is only possible when a cell has three alive neighbours, as shown in Fig.~\ref{fig:Fig2}. For any number of neighbours different to three, the shift in the S-shaped curve has raised the threshold population needed to switch to the higher intensity state, such that the non-resonant pulse is insufficient.

While our scheme may appear complicated, involving different components of the coherent driving field $F_+$ and $F_-$ as well as components of the non-resonant pulse $P_+$ and $P_-$, we note that these just correspond to specific polarizations of a continuous wave laser and a pulse. Both can be spatially uniform with no specific site to site modulation. Such pulse induced switching of a system driven by a continuous wave driving field is well within the limits of current technology~\cite{Adrados2011,Anton2012,DeGiorgi2012,Ohadi2015,Cerna2013}. Each automaton update requires the application of just one pulse (see sec. \ref{section:Pulse sequence}).
%

\section*{Conway's life and other cellular automata}
\label{section:Conway's life and other cellular automata}
We varied the parameters $P_+$ and $P_-$ and found that different automata rules were possible. For example, the rule where a cell is born if it has three neighbours and survives if and only if it has less than seven neighbours alive was possible~(see sec. \ref{section:Other cellular automata}). This specific rule has not appeared in the literature but is also universally complete~(see sec. \ref{section:Other cellular automata}). More importantly, the presence of this rule shows that it is possible to realize the behaviour of overpopulation, where a cell dies if too many of its neighbours are alive. This allows more complex automata.

Conway's life requires that a cell is born if it has three neighbours and survives if and only if it has two or three neighbours alive. Fig.~\ref{fig:Fig6} shows the numerically calculated probability for a cell to be born or to survive if it has a specific number of neighbours alive, where we have adjusted the field $F_-$ to $F_-=-J\left(2.8\psi_{\rm{alive}}+5.2\psi_{\rm{dead}}\right)$. This gives a slight bias to allow the cell with two alive neighbours to survive and we find that for larger pulse intensity Conway's life is obtained (see, e.g., the red marker in Fig.~\ref{fig:Fig6}). Fig.~\ref{fig:Fig8} shows the dynamics under a non-resonant pulsed excitation for different numbers of neighbours, confirming that Conway's life is attained in the system. In~sec. \ref{section:Effect of same polarization coupling} we have also shown that Conway's life can be achieved even in the presence of an additional polarization conserving coupling between neighbours.

\section*{Solitons and Patterns}
\label{section:Solitons and Patterns}
Having established the presence of complex cellular automata, we show examples of the structures that can be formed in the polariton system. Conway's life is known to support stable self-localized states (Fig.~\ref{fig:Fig4}(a)) in analogy to the solitons considered in Ref.~\cite{Ostrovskaya2013}. As we operate on a lattice, these can be considered as discrete solitons~\cite{Egorov2013}, which are also dissipative~\cite{Ostrovskaya2012}. These solitons are permanent in the system so long as the near-resonant driving field is applied; they do not decay even accounting for the finite polariton lifetime. Another example is shown in Fig.\ref{fig:Fig4}(bi,bii), which illustrates a rotating soliton. In addition, since bistability at each lattice site exists for some range of detuning ($\Delta$), these solitons and the cellular automaton behaviour in general survives in the presence of disorder~(see sec. \ref{section:Effect of disorder}).

Conway's life also supports propagating oscillating solitons, known as gliders or spaceships (see Fig.~\ref{fig:Fig4}(ci-cv)), which can be generated in a train using a glider gun (Fig.~\ref{fig:Fig4}(di,dii)). Collisions between gliders give rise to the formation of a very wide variety of intricate structures, such as exploding patterns (Fig.~\ref{fig:Fig4}(ei,eii), cf.~\cite{Xue2014}), puffer trains (Fig.~\ref{fig:Fig4}(fi,fii), cf.~\cite{Skryabin2017}), and the possibility of ordered or chaotic patterns (see sec. \ref{section:Chaos  in cellular automata}, c.f.~\cite{Gavrilov2018}). The proof of universal completeness of cellular automata can be established considering signals carried by ``ladders''~(see sec. \ref{section:Other cellular automata}). The sensitivity of our result in Fig.~\ref{fig:Fig6} shows that stable lasers with less than $1\%$ intensity fluctuation will be required. We note that intensity fluctuations on the scale of $0.02\%$ are used in polariton experiments~\cite{Abbaspour2015}.

\section*{Conclusion}
We predict that when a particular stimulus is applied to a polariton lattice, namely a series of identical non-resonant excitation pulses, artificial life can appear in the form of a cellular automaton. This allows the realization of a variety of fundamental nonlinear optical structures under the same conditions, such as  solitons, oscillating solitons, breathers, and various patterns. The complexity that arises from the combination of these phenomena is encapsulated within the simple update rules of the automaton. This further shows that polariton solitons are universally complete, even in the presence of dissipation and disorder.


\section*{Acknowledgments}
This work was supported by the Ministry of Education (Singapore) grant 2017-T2-1-001.

\section{Other cellular automata}
\label{section:Other cellular automata}
We varied  $P_{+}$ and $P_{-}$ for $F_-=-J\left(3\psi_{\rm{alive}}+5\psi_{\rm{dead}}\right)$ and plotted the probability of a cell to be alive or dead depending on their neighbours as shown in Fig.~\ref{fig:Fig3}. We found different rules like the life without death (B3/S012345678, using standard notation of automata specifying the number of neighbours needed for a cell to be born/survive) and the rule (B3/S0123456) where a cell is born if it has three alive neighbours and survives if it has less than seven alive neighbours. In the case of the Life without death automaton, we take the parameters corresponding to the red spot in Fig.~\ref{fig:Fig3} while the green spot indicates the parameters for which we achieve the rule B3/S0123456.\par

To show that the rule B3/S0123456 is universally complete, we follow Ref.~\cite{GriffeathMoore}, which previously showed that the Life without death automaton is universally complete. The proof was based on considering signals in the form of repeating structures, known as ``ladders'', that grow from a finite seed (Fig.~\ref{fig:Fig7}(ai)) in a particular direction as shown in Fig.~\ref{fig:Fig7}(aii). If ladders can be terminated, turned, and blocked by other ladders then they can be suitably combined into logic gates (``OR",``AND", ``NOT") to realize universal completeness~\cite{GriffeathMoore}. These situations are demonstrated in Figs.~\ref{fig:Fig7}(bi-dii) for the rule B3/S0123456.\par

It should be noted that when one ladder is used to block another, the relative position of ladders is important. If ladders have the incorrect relative position, then their collision creates extra unnecessary fluctuations in the system, called ``lava"~\cite{GriffeathMoore}, which may divert the propagation of the ladder into a different direction. As the horizontal periodicity of the considered ladders is four and the horizontal ladder can collide with an occupied or unoccupied portion of the vertical ladder, there are a total of eight possible relative shifts denoted by ($h,v$), where $h \in \Z_{4}$ and $v \in \Z_{2}$ are the horizontal and vertical shifts respectively (note that in Ref.~\cite{GriffeathMoore} these shifts are called phase shifts, but we avoid this terminology here so as to not confuse with an optical phase). In Fig.~\ref{fig:Fig7}(di), two ladders incident from the left have phases (0,0) and (2,1) and are blocked by the ladder on the right. In Fig.~\ref{fig:Fig7}(ei), the horizontal ladder will collide with the vertical ladder at the unoccupied portion of the vertical ladder, which will not result in the correct blocking behaviour. However, ladders can also be shifted using particular seeds. For example, the situation in Fig.~\ref{fig:Fig7}(eii), causes the approaching ladder to shift in space so as to collide with the correct relative position with respect to the blocking ladder as shown in Fig.~\ref{fig:Fig7}(eiii). In this way clean blocking can always be arranged by shifting ladders appropriately between collisions. \\

\section{Chaos  in cellular automata}
\label{section:Chaos  in cellular automata}
So far we have seen that different patterns emerge from Conway's life, including stable, periodic, and moving patterns. Here we provide an example of chaotic behaviour. We start with the initial condition shown in Fig.~\ref{fig:Fig5}(ai), which gives rise to Fig.~\ref{fig:Fig5}(aii) after 59 time steps. Changing the initial condition by just one cell to that of Fig.~\ref{fig:Fig5}(bi), we arrive at a completely different pattern (see Fig.~\ref{fig:Fig5}(bii)) after the same number of iterations.\\

\section{Materials}
\label{section:Materials}
Two-dimensional polariton lattices have been intensely developed using GaAs \cite{Kim2011,Winkler2015,Sala2015,Whittaker2018}, so these are currently the most appropriate materials for our proposal. Of course real systems also come with an inevitable disorder, which in principle can destroy polariton solitons. In our system, while the stationary states plotted in Fig.~\ref{fig:Fig1}(c) can be shifted in the presence of disorder, it is notable that bistability survives over some range of detunings, $\Delta$. Furthermore, introducing a distribution of values of $\Delta$ varying from site-to-site we found that cellular automaton rules could persist provided that the strength of disorder is below a finite threshold, estimated around 30 $\mu eV$ (see sec. \ref{section:Effect of disorder}) in GaAs, which is comparable to the disorder of state-of-the-art lattices \cite{Baboux2016}. As our scheme is generic it could also be compatible with materials capable of operation at higher temperatures. For example, Te and Se-based micropillars are in development and could be assembled in the same way as their GaAs predecessors \cite{KleinKlembt2015, RoussetaPietka2015}.  \\

\section{Effect of disorder}
\label{section:Effect of disorder}
In Fig.~\ref{fig:Fig9} we calculated the success rate in the presence of disorder, represented by variation in the detuning $\Delta$ at different lattice sites. The disorder in $\Delta$ is taken with a uniform distribution with peak-to-peak magnitude $\delta\Delta$. It is notable that when the disorder strength is below some limiting threshold, different automaton rules are obtained with $100\%$ accuracy. Even above the limiting threshold the success rate remains high. It should also be noted that these curves only represent a lower bound on the success rate for a given task. If we consider the on-site disorder of the micropillar of the order of $30~ \mu$eV, then with polariton lifetime $1.8~$ps, the disorder is approximately 0.09 in a dimensionless unit (see the gray vertical line in Fig.~\ref{fig:Fig9} ). Even if some configurations of an automaton do not update with $100\%$ accuracy, the configurations that do update correctly may be sufficient for a particular task, e.g., formation of ladder, creation of a propagating soliton  or implementation of an imaging processing algorithm. We have separately verified that in the presence of disorder at the level of 0.09 in our dimensionless units all the structures required for universal computation are perfectly maintained.\\

\section{Pulse sequence}
\label{section:Pulse sequence}
As mentioned earlier, one non-resonant pulse of the form $P_{\pm}(t)=P_{\pm}e^{- \Gamma_R t}$ is needed for each automaton update. As an example we show how an oscillating soliton-blinker updates to the next state after applying  the pulse. Initially we start with the configuration given by  Fig.~\ref{fig:pulse_seq}(j) and solve Eq.~\ref{eq:GP} with $F_{+}$ along with $F_-=-J(2.8\psi_\mathrm{alive}+5.2\psi_\mathrm{dead})$ for time unit, $t=(0,20)$. Then we apply the  non-resonant pulse  to all lattice sites (whose amplitude is indicated by red dot in  Fig.~\ref{fig:Fig6}) as shown by dotted curves in Fig.~\ref{fig:pulse_seq}. The steady state updates to Fig.~\ref{fig:pulse_seq}(k), which we expect if the system follows Conway's life as depicted in Fig.~\ref{fig:Fig4}(bi-bii).\\

\section{Effect of same polarization coupling}
\label{section:Effect of same polarization coupling}
To check the effect of the same polarization coupling with the neighbours we add a term $J_P\sum_{\langle m\rangle}\psi_{m,\pm}$ with Eq.~\ref{eq:GP} such that the dynamics of the micropillars are governed by the following equation 
\begin{align}
i\frac{\partial\psi_{n\pm}}{\partial t}&=\left(-\Delta+|\psi_{n\pm}|^2+\alpha_2|\psi_{n\mp}|^2+iP_\pm(t)-\frac{i}{2}\right)\psi_{n\pm}+J_P\sum_{\langle m\rangle}\psi_{m,\pm}+J\sum_{\langle m\rangle}\psi_{m,\mp}+F_\pm.\label{eq:JGP}
\end{align}
Next we perform all the steps as described earlier. Indeed, the Conways life can be achieved by choosing  proper values of the incoherent pulses (see Fig.~\ref{fig:TB_Scan_JP}). In the calculation we have chosen $J_P=J=0.01$, however, we have checked that the Conways life still exists for $J_P>J$ as long as $J_P$ is small ($J_P\ll1$).\\

\section{Continuous model}
\label{section:Continuous model}
\begin{figure}[htb]

\includegraphics[width=0.7\linewidth]{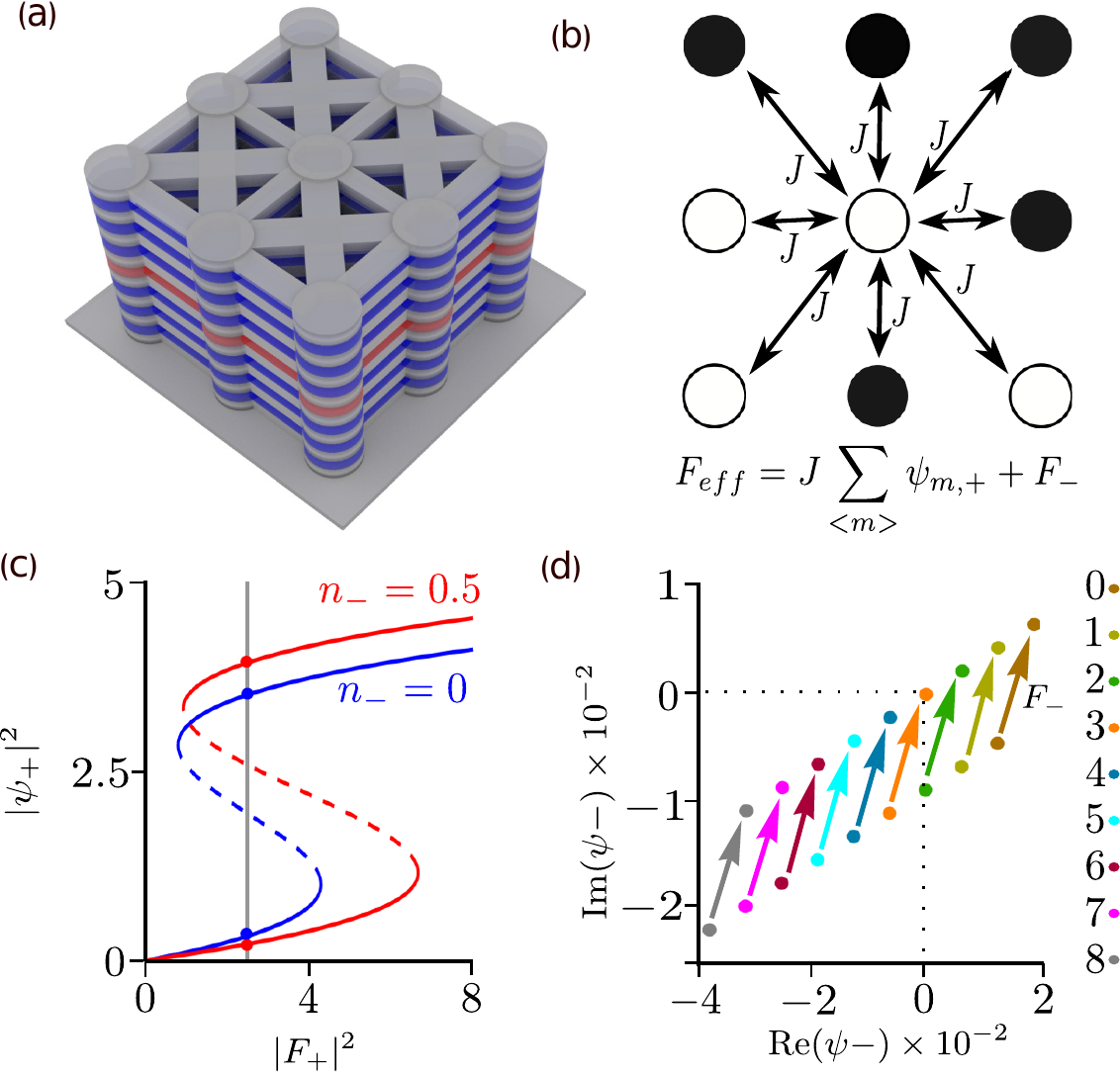}
\centering
\caption[lipsum2]{\textbf{(a)} A square lattice of coupled cells formed by polariton resonators. Blue and grey denote alternating layers of refractive index, which trap polaritons in the cavity layer (red). The sample is etched in a way to leave channels that connect the resonators. \textbf{(b)} Each cell interacts with eight neighbours, which provide an effective driving, $F_\mathrm{eff}$. This field depends on how many neighbours are alive and on the constant field $F_{-}$. \textbf{(c)} Single cell stationary solutions under the condition  $J=0, P_{\pm}(t)=0$ , and $ F-=0$ with and without the influence of $\psi_-$. A non-zero population of $\sigma_-$ particles raises the threshold intensity for reaching the higher $\psi_+$ state. Parameters: $\Delta=3, \alpha_2=-1$ \cite{VladimirovaCronenberger2010}. The grey line shows the fixed continuous driving for which there are two stable solutions (dashed curves correspond to unstable states). \textbf{(d)} Neighbour dependent values of the $\psi_-$ field caused by the effective driving, $F_\mathrm{eff}$, with (upper points) and without (lower points) the driving field $F_{-}$ under the condition $P_{\pm}(t)=0$ for a cell initially in the lower intensity (dead) $\psi_+$ state. Colours represent different numbers of neighbours in the higher intensity (alive) $\psi_+$ state. Parameters: $J=0.01$, $\alpha_2=-1$ \cite{VladimirovaCronenberger2010}.}
\label{fig:Fig1}
\end{figure}

\begin{figure}[htb]
\includegraphics[width=0.7\linewidth]{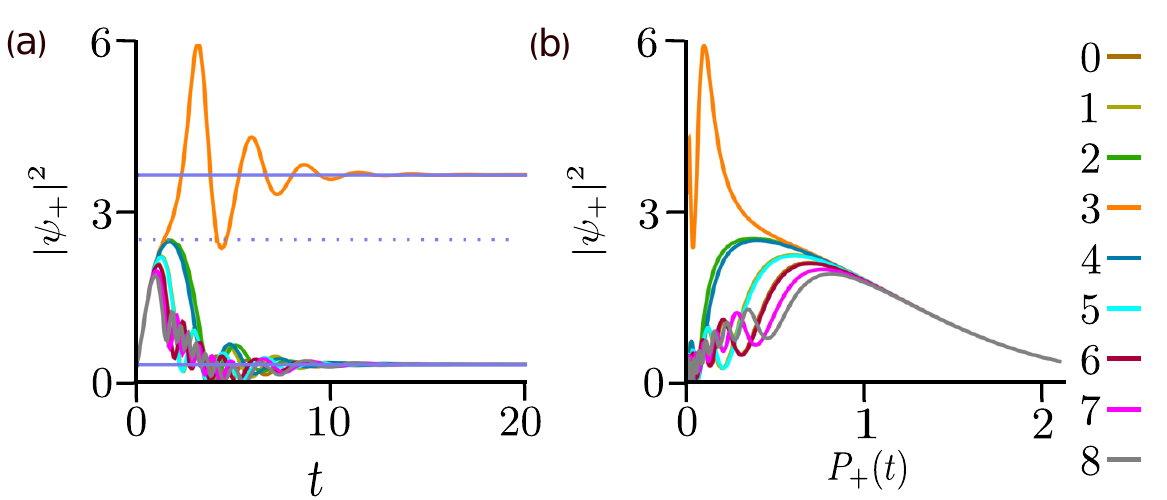}
\centering
\caption{\textbf{(a)} Time evolution of $\psi_{+}$ after the application of a non-resonant pulse. The  solid lines (violet color) represent the stable states. The dotted line represents the unstable state. The figure shows a switching to the higher intensity state if a dead cell has three neighbours alive, while otherwise it remains in the lower intensity state. \textbf{(b)} Plot of the $\sigma_+$ intensity vs the decaying non-resonant pulse intensity for different numbers of alive neighbours. Parameters: $\Gamma_{R}=1$.}
\label{fig:Fig2}
\end{figure}

\begin{figure}[htb]
\includegraphics[width=0.5\linewidth]{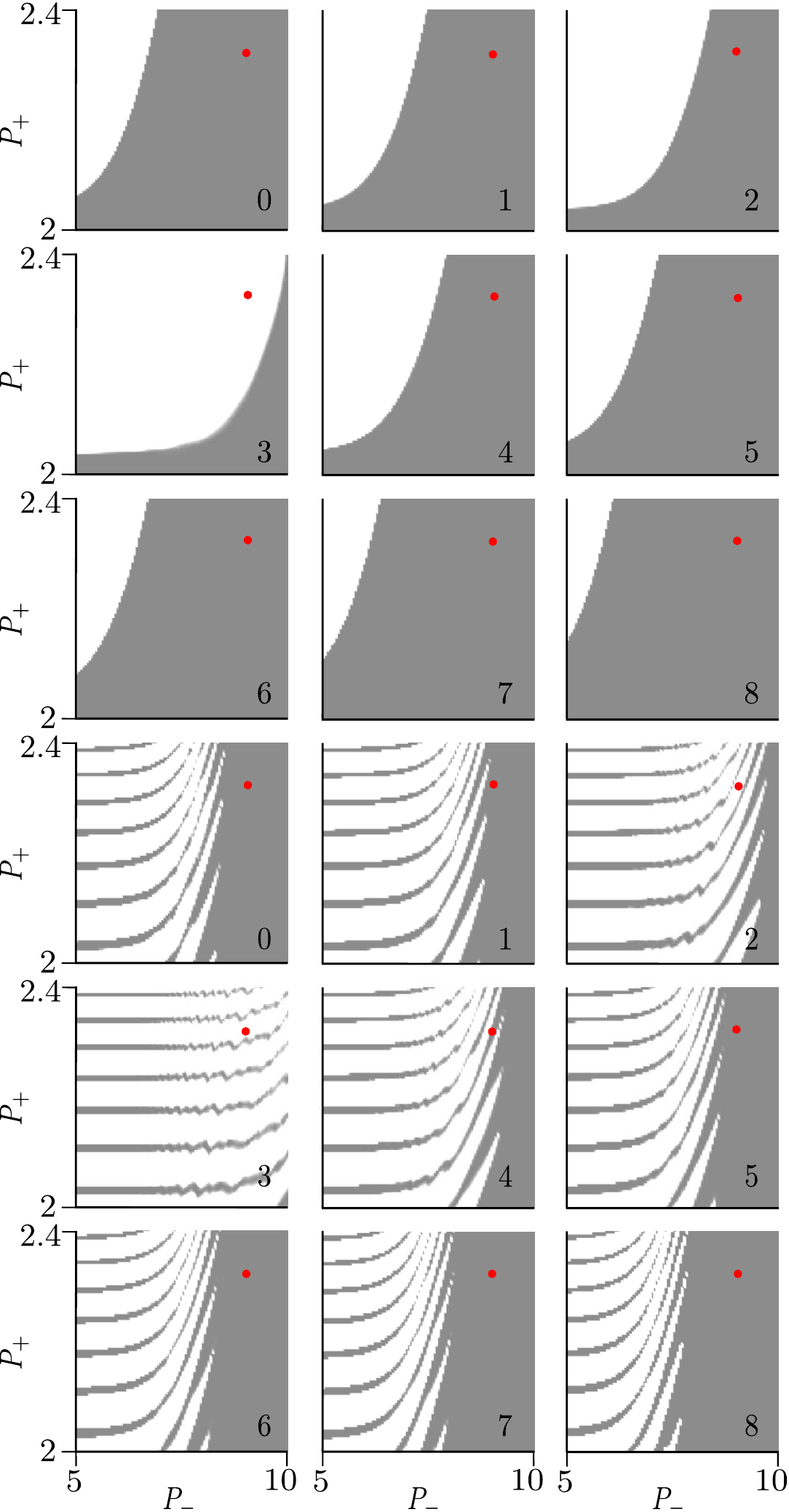}
\centering
\caption{Variation of automaton rules with $P_+$ and $P_-$. Each plot shows whether a cell that is initially  dead (upper nine plots) or initially alive (lower nine plots) finishes in an alive state after the application of a pulse for different numbers of neighbouring (initially) alive cells (given by the labels in the bottom-right corners of each plot). White corresponds to parameters for which a cell finishes in the alive state, while grey corresponds to parameters for which a cell finishes in the dead state. The red spot indicates a parameter choice for which Conway's life appears. Parameters: $F_-=-J(2.8\psi_\mathrm{alive}+5.2\psi_\mathrm{dead})$, $\Delta=3,J=0.01$, and $\alpha_2=-1$.}
\label{fig:Fig6}
\end{figure}

\begin{figure}[htb]
\includegraphics[width=0.7\linewidth]{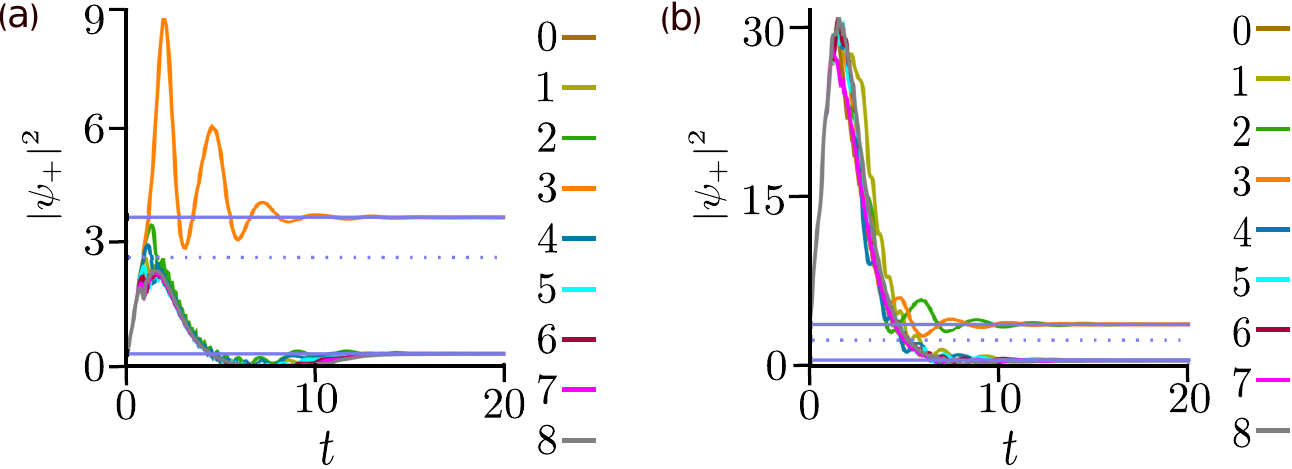}
\centering
\caption{Dynamics of $|\psi_+|^2$ under pulsed excitation for a cell initially  dead \textbf{(a)} or alive \textbf{(b)}. The different coloured curves correspond to different numbers of neighbours (initially) alive. Only configurations where a dead cell has three alive nieghbours, or where a cell was initially alive and has two or three alive neighbours results in the cell being alive at the end of the pulse. The amplitude of the pulses  were chosen according to the red spot in Fig.~\ref{fig:Fig6}. Parameters: $F_-=-J(2.8\psi_\mathrm{alive}+5.2\psi_\mathrm{dead})$, $\Delta=3,J=0.01$, and $\alpha_2=-1$.}
\label{fig:Fig8}
\end{figure}

\begin{figure}[htb]
\includegraphics[width=0.5\linewidth]{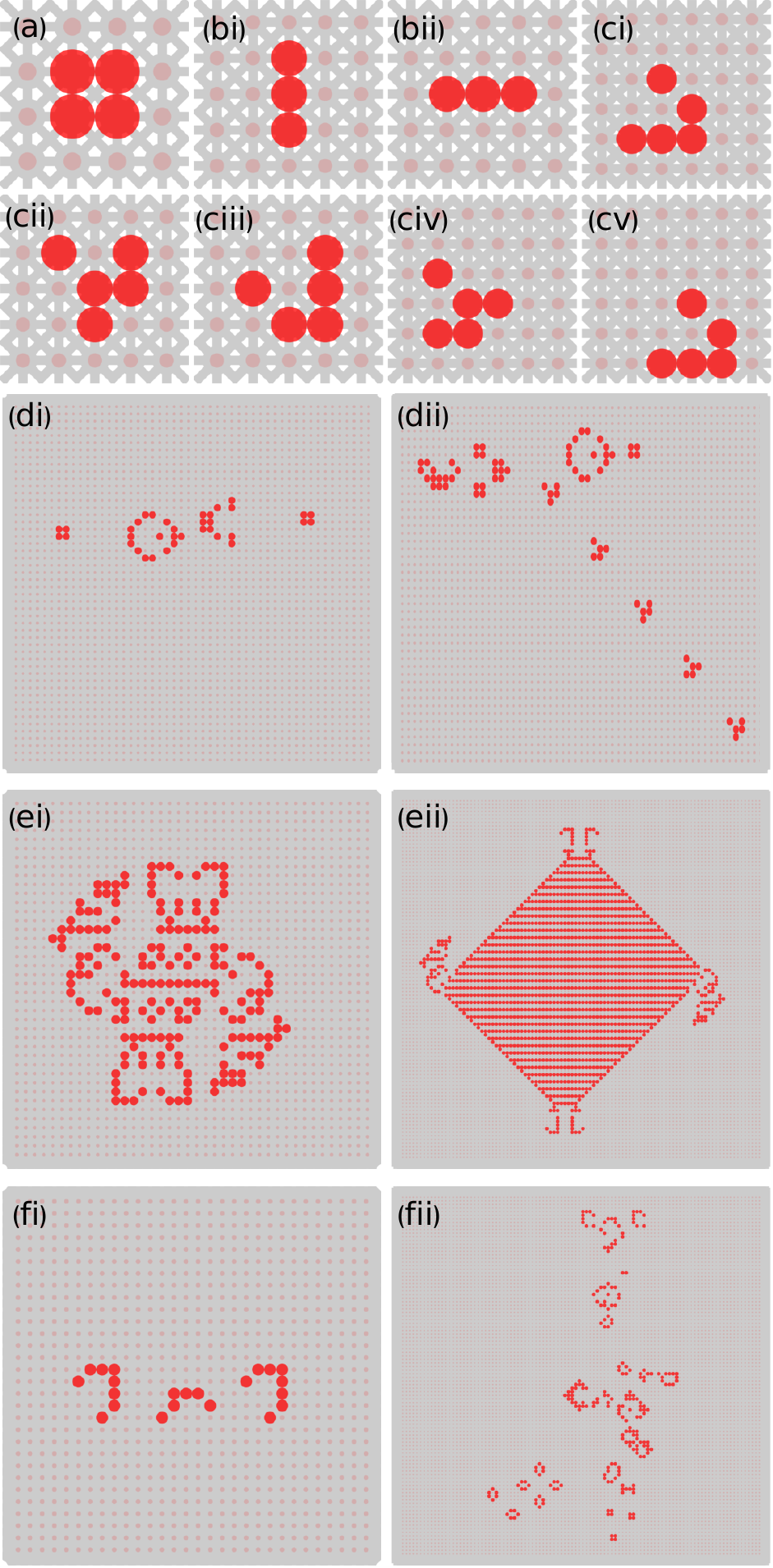}
\centering
\caption{Examples of configurations attained within Conway's life. \textbf{(a)} A $2\times2$ block maintains its form over time. \textbf{(bi-bii)} An oscillating soliton (blinker), with periodicity two. \textbf{(ci-cv)} A glider, that is, a soliton that translates diagonally in four timesteps. \textbf{(di,dii)} A glider gun, that is, starting with the configuration in \textbf{(di)}, a train of solitons is emitted as shown in \textbf{(dii)}. \textbf{(ei,eii)} A soliton explosion where an initial condition \textbf{(ei)} fills the whole space over time \textbf{(eii)}. \textbf{(fi,fii)} An emitting soliton where the initial condition in \textbf{(fi)} moves upwards, while also leaving behind a dynamic pattern.}
\label{fig:Fig4}
\end{figure}

\begin{figure}[htb]
\includegraphics[width=0.55\linewidth]{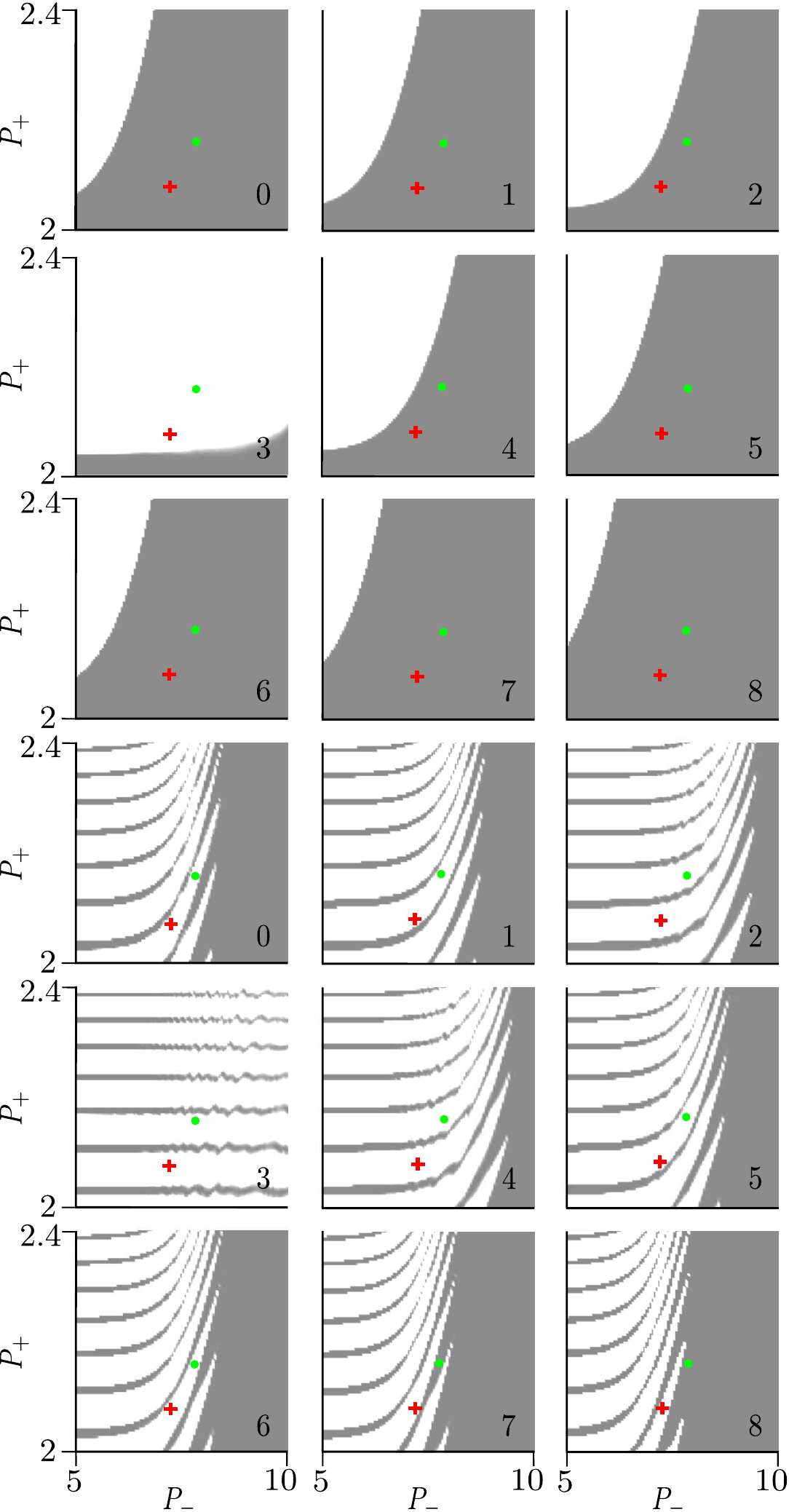}
\centering
\caption{\textbf{(a)} Probability for a cell to be in the alive state depending on the number of alive neighbours (marked in the bottom right corner of the plots) for $F_-=-J\left(3\psi_{\rm{alive}}+5\psi_{\rm{dead}}\right)$, for cases when it was initially dead (upper nine plots) and alive (lower nine plots). For the parameters in the white region, the cell finishes in the alive state and for the grey region, it finishes in the dead state. The green dot marks a set of parameters for which the rule B3/S0123456 is obtained, while the red + marks a set of parameters for which Life without death (B3/S012345678) was obtained.}
\label{fig:Fig3}
\end{figure}

\begin{figure}[htb]
\centering
\includegraphics[width=0.3\textwidth,height=17 cm]{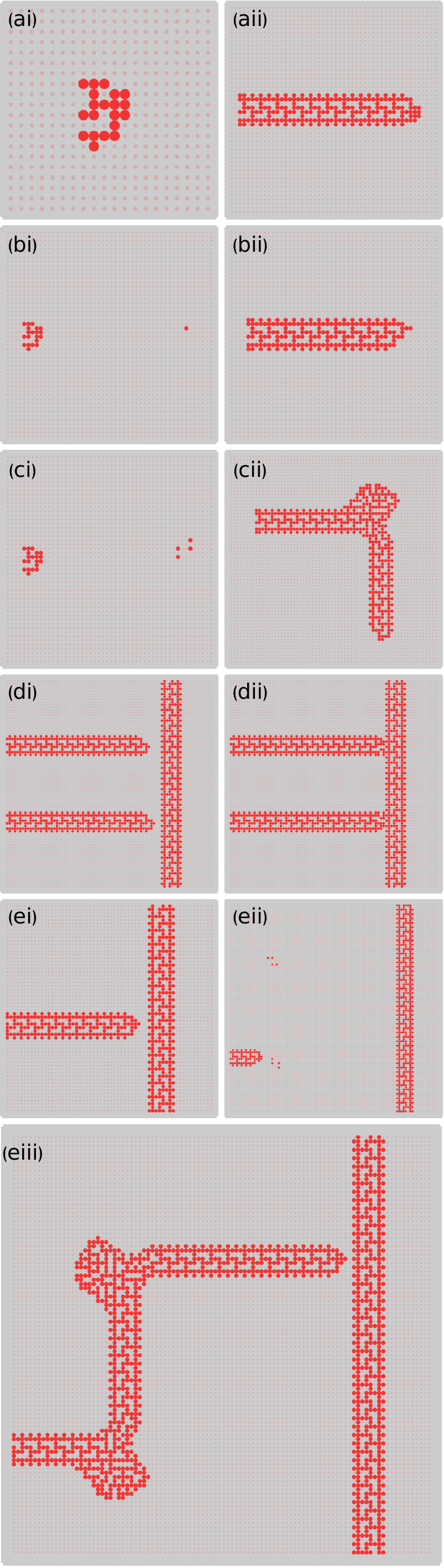}
\caption{Behaviour of ladders emerging from the rule (B3/S0123456). \textbf{(ai)} It is a finite seed that gives a periodic structure that moves in one direction, called a ladder shown in \textbf{(aii)}.  \textbf{(bi,bii)} The propagation of a ladder is ended by a single cell. \textbf{(ci,cii)} In the presence of a finite seed a ladder is turned so as to propagate in the perpendicular direction. \textbf{(di,dii)} Ladder can be blocked by other ladders. Here two ladders are incident from the left with relative phases (0,0) and (2,1).  \textbf{(ei-eiii)} The relative displacement of ladders can be shifted, which helps to ensure clean blocking.}
\label{fig:Fig7}
\end{figure}

\begin{figure}[htb]
\includegraphics[width=0.5\linewidth]{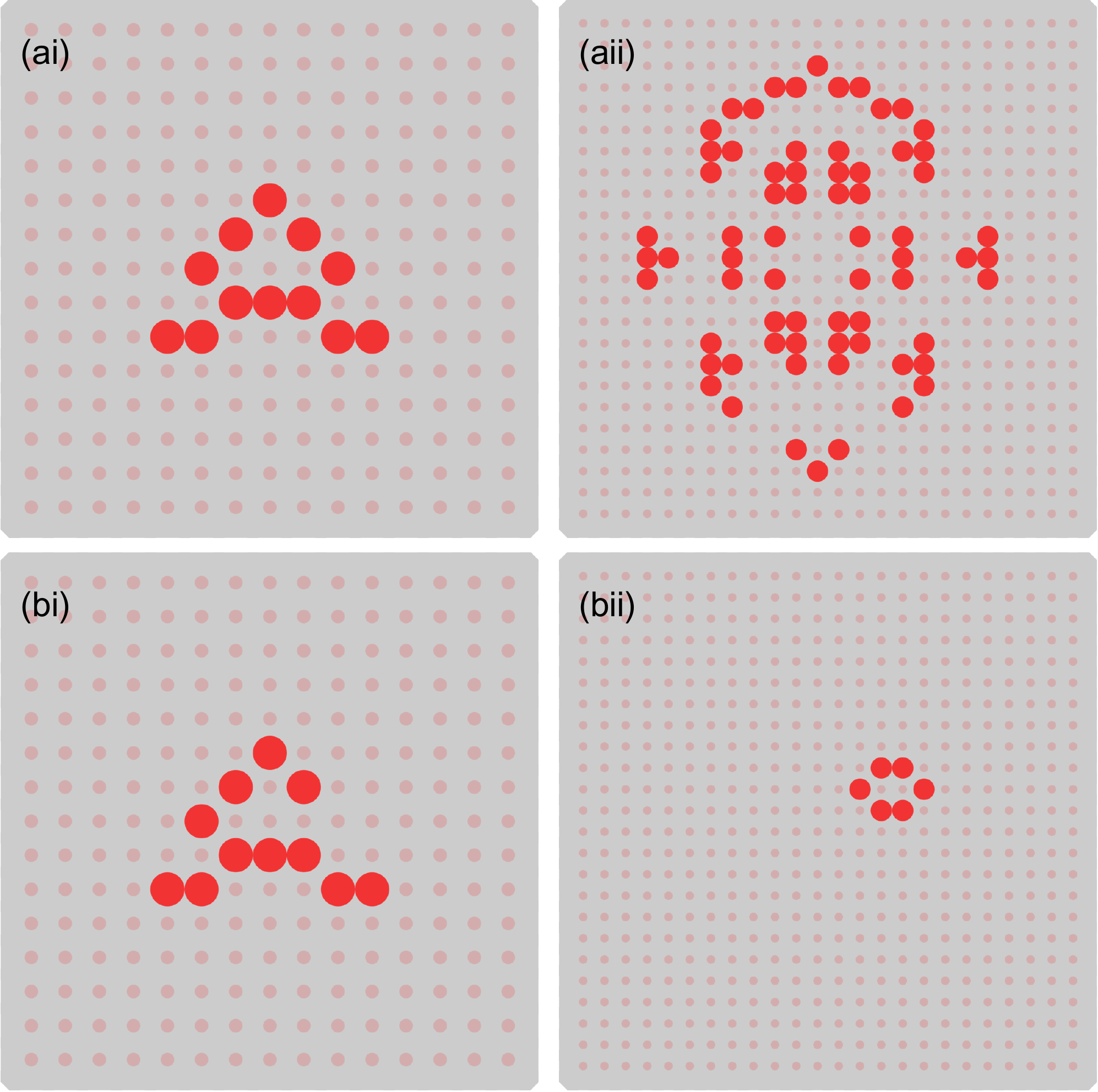}
\centering
\caption{Example of chaotic behaviour in Conway's life. \textbf{(ai,aii)} The initial state \textbf{(ai)} evolves into the pattern \textbf{(aii)} after 59 iterations. \textbf{(bi,bii)} The initial state \textbf{(bi)} differs from that of \textbf{(ai)} by only one cell, and gives rise to a completely different pattern \textbf{(bii)} after the same number of iterations.}
\label{fig:Fig5}
\end{figure}

\begin{figure}[htb]
\includegraphics[width=0.5\linewidth]{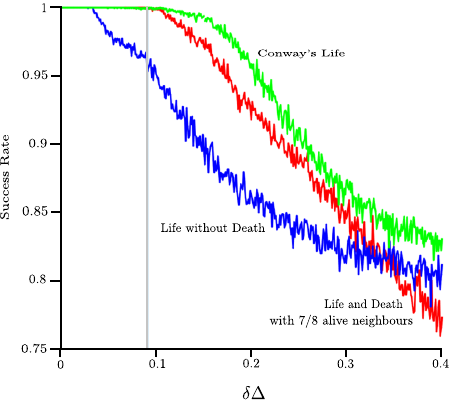}
\centering
\caption{Dependence of the success rate on disorder for different cellular automaton rules. $\Delta$ is chosen differently for different lattice sites, using a uniformly distributed random variable with peak-to-peak amplitude $\delta\Delta$. When disorder is below some threshold, cellular automata are still obtained perfectly with $100\%$ success rate. Above this disorder threshold, the success rate slowly decreases.}
\label{fig:Fig9}
\end{figure}

\begin{figure}[htb]
\includegraphics[width=0.8\linewidth]{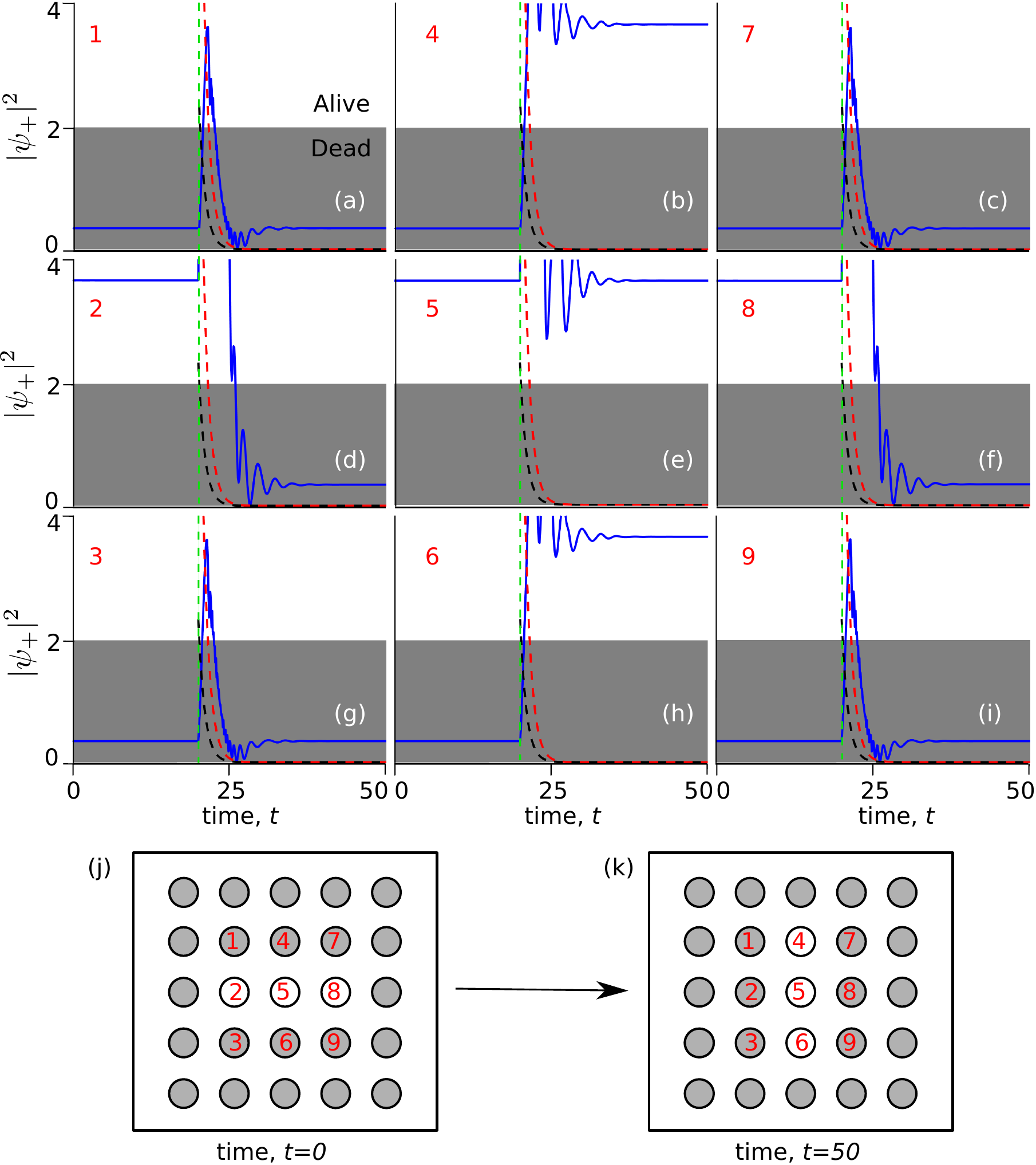}
\centering
\caption{ Time evolution of oscillating soliton - blinker which emerges from Conway's life. After applying the pulse the initial state (j) updates to final state (k). The grey region  corresponds to a dead state while white to an alive state. Here we plotted the time dynamics only for the centered $3\times 3$ lattice sites indicated by indices $1, 2, 3..., 9$ (top-left). It shows that initially we start with an alive state in cells  $(2,5,8)$ and after the pulse two new cells are born ($4$ and $6$) and $5$ remains alive. The pulses are indicated by the dotted curves where black represents $P_+(t)$ and red represents $P_-(t)$. The green dotted vertical lines represent the time at which the pulses are injected. The amplitude of the pulses is taken from the red dot in Fig. \ref{fig:Fig6}.}
\label{fig:pulse_seq}
\end{figure}

\begin{figure}[htb]
\includegraphics[width=0.6\linewidth]{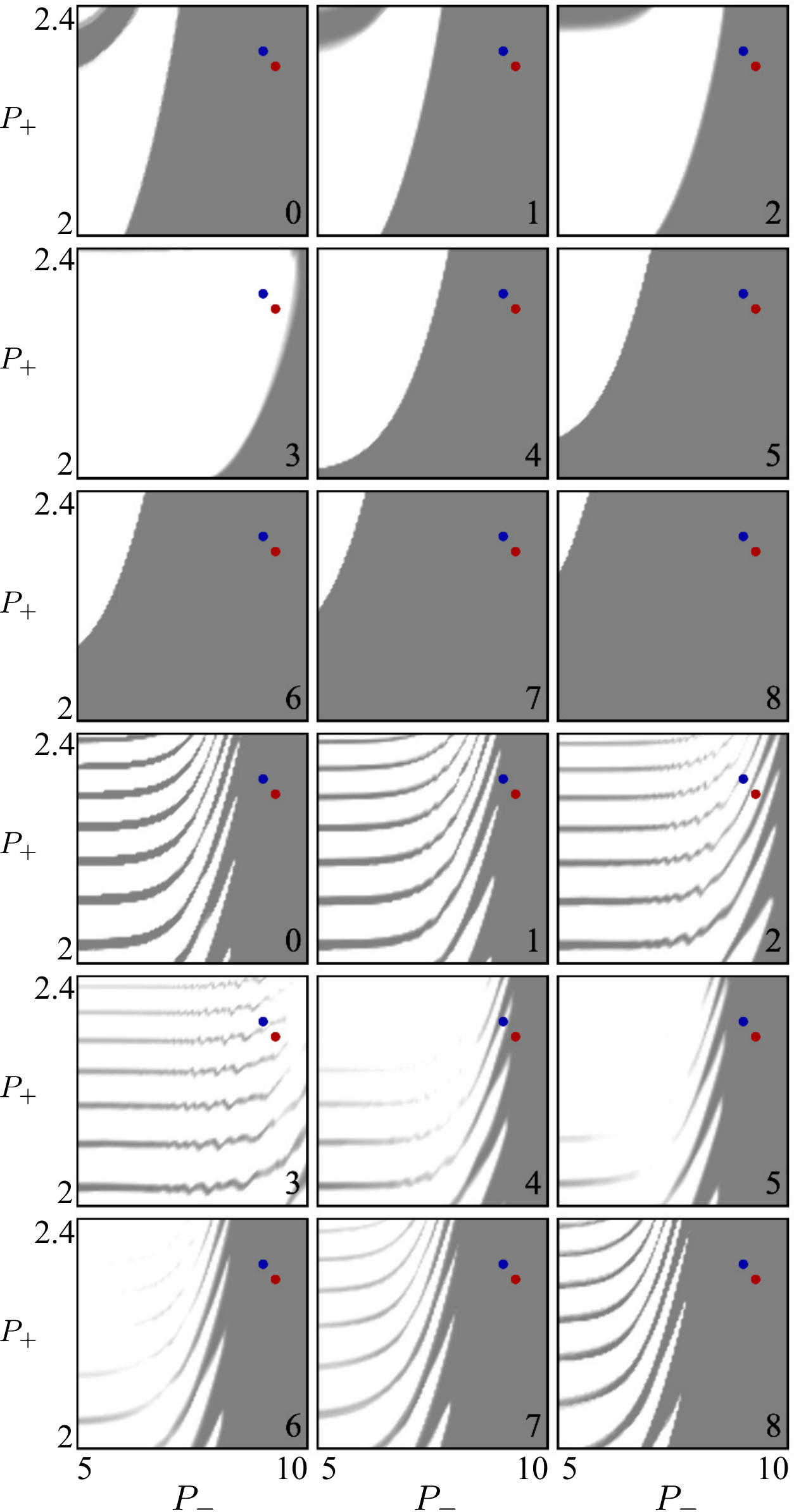}
\centering
\caption{Variation of automaton rules with $P_+$ and $P_-$ in the tight binding model according to Eq.~(\ref{eq:JGP}) where 
polarization conserving coupling with neighbours are also considered. The parameter indicated by the red dot corresponds to the Conways life. The blue dot corresponds to the parameter corresponding to the Conways life shown in Fig~\ref{fig:Fig6} where $J_P=0$. Parameters: $J_P=J=0.01$, $F_-=-J\left(2.8\psi_{\rm{alive}}+5.2\psi_{\rm{dead}}\right)$ and all other parameters are kept the same as those in Fig.~\ref{fig:Fig6}.}
\label{fig:TB_Scan_JP}
\end{figure}

\begin{figure}[htb]
\includegraphics[width=0.8\linewidth]{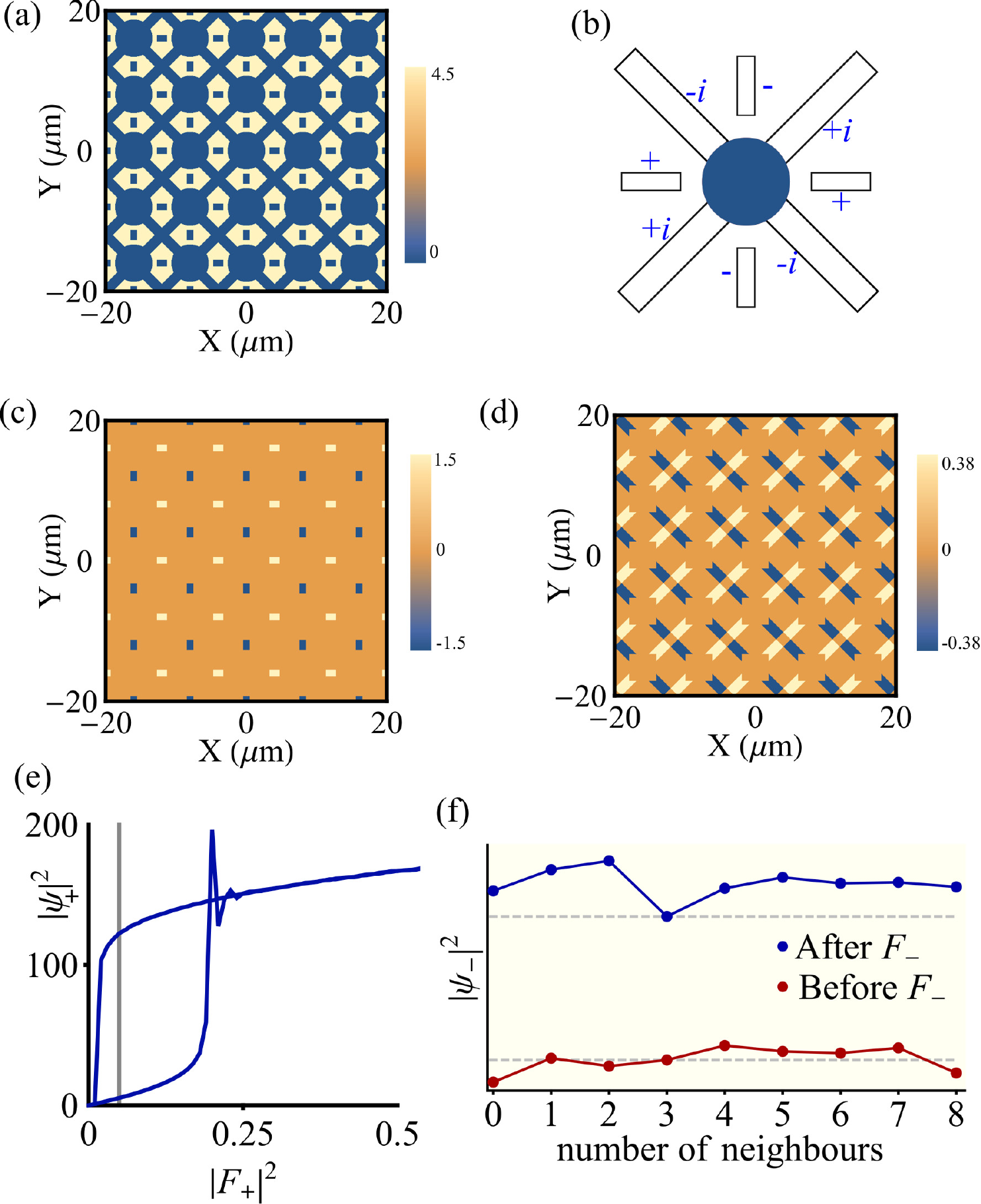}
\centering
\caption{(a) An example of the potential profile of a 5$\times$5 lattice. Each micropillar is connected to its eight neighbours through the channels. (b) Due to their orientations the horizontal and vertical channels have opposite polarization splittings. The polarization splitting inside the diagonal channels have only imaginary parts due to their $\pi/4$ orientation with respect to the $x$ axis and those inside the antidiagonal channels are opposite to the diagonal ones. (c-d) The real and imaginary parts of the polarization splitting, respectively. (e) The bistability curve of the micropillars  where the gray horizontal line indicates the $|F_+|^2$ value used in the calculation. (f) The effect of $F_-$ on the initially dead configurations with different number of alive neighbours. After applying $F_-$ the configuration corresponding to three alive neighbours has the minimum $|\psi_-|^2$, similar to Fig.~\ref{fig:Fig1}(d).}
\label{fig:conti_pot_bi}
\end{figure}

\begin{figure}[htb]
\includegraphics[width=0.6\linewidth]{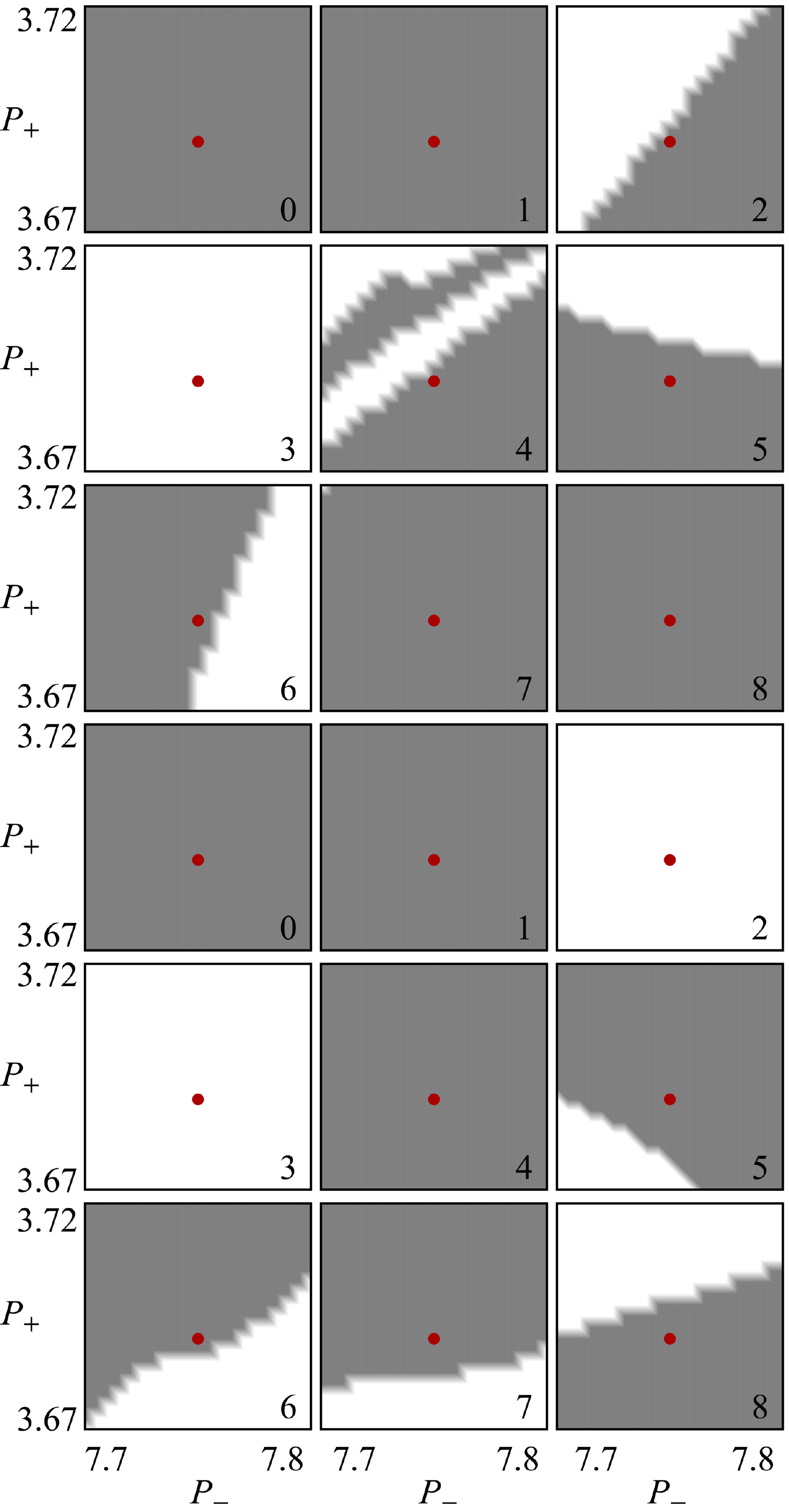}
\centering
\caption{ Variation of automaton rules with $P_+$ and $P_-$ in the continuous model under the periodic boundary condition is depicted here. White corresponds to parameters for which a cell finishes in an alive state, while grey corresponds to parameters for which a cell finishes in a dead state. Corresponding to the nine upper  plots, the states were initially dead and alive for the lower nine plots. The red dot indicates a parameter choice for which Conway's life appears. Parameter: $\alpha_2=-1$. }
\label{fig:conti_pscan}
\end{figure}

\begin{figure}[htb]
\includegraphics[width=0.4\linewidth]{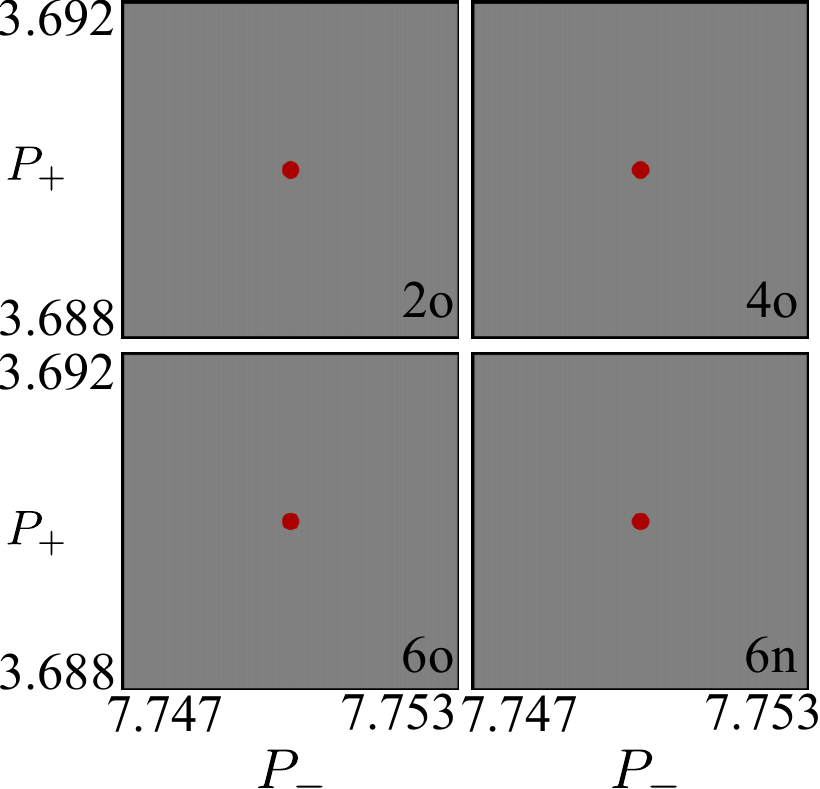}
\centering
\caption{ Zoomed view of the 2o, 4o, 6o and 6n configurations, respectively, from the previous figure which show that the chosen parameter can tolerate the 0.02$\%$ variation of the incoherent pulses. Here `o' and `n'  indicate initially dead and alive states, respectively.}
\label{fig:Zoomed_conti_pscan}
\end{figure}

In this section we model the system with a continuous wave-function to represent the coupled exciton-polariton micropillars shown in Fig.~\ref{fig:conti_pot_bi}(a). The diameter of the micropillars is chosen to be $5.3 ~\mu$m \cite{BoulierBamba2014} and the separation between the micropillars is 8 $\mu$m along the vertical and horizontal directions. To realize the diagonal  couplings with the same strength as the horizontal and vertical ones we take the diagonal channels wider. Such a structure with channel connecting two pillars was studied in Refs.~\cite{Bayer1998,Rui_2020}. The width of the channels along the vertical and horizontal direction is taken as 0.5 $\mu$m and those along the diagonal and anti-diagonal direction is taken as 2 $\mu$m. The depths of the micropillars as well as the channels are taken to be around  $4.5$ meV. The  polariton mass and lifetime are taken as  $3 \times 10^{-5} m_{e}$, where $m_{e}$ is the free electron mass, and $3$ ps, respectively. Due to the shape anisotropy of the channels a splitting between modes polarized parallel and perpendicular to the channel naturally occurs inside them. We take the polarization splitting inside channels as $V_T=V_{avg}/\Delta L$, where $V_{avg}$ is the average splitting in the system which is taken as 0.75 meV $\mu$m and $\Delta L$ is the width of the respective channels. Since different channels have different orientations, a phase factor $\exp(2i\phi)$ appears in $V_T$, where $\phi$ is the angle of the channels, to ensure that the polarization splitting is always in the correct direction (this phase factor is readily derived from first writing a splitting between $x$ and $y$ directions and performing a rotation). The phase factors are shown in Fig.~\ref{fig:conti_pot_bi}(b). Due to the symmetry of the unit cell it is clear that there will be no  polarization splitting
inside the pillars. The spatial profiles of the real and imaginary parts of the  polarization splitting is shown in \ref{fig:conti_pot_bi}(c-d), respectively. Now that all the ingredients are ready, we start by solving the following driven-dissipative Gross-Pitaevskii  equation: 
\begin{align}
i\hbar\frac{\partial\psi_{\pm} (\vec{x})}{\partial t}=&\Big(-\frac{\hbar^{2}}{2m}\left(\frac{\partial^2}{\partial x^2}+\frac{\partial^2}{\partial y^2}\right)+V(\vec{x})-\Omega_0-\frac{i \Gamma}{2}+\alpha_1 |\psi_{\pm} (\vec{x})|^2+\alpha_2 |\psi_{\mp} (\vec{x})|^2 \nonumber\\
& +iP_{\pm}(t)\Big) \psi_{\pm}+V_{T}(\vec{x},\phi)\Psi_{\mp}+F_{\pm}.
\label{eq:sscheq}
\end{align}
Here $V(\vec{x})$ is the potential profile represented in Fig.~\ref{fig:conti_pot_bi}(a) and $\Omega_0$ is the energy of the resonant pump, which we take $1.1$ meV.  To find the suitable parameters easily, we move to dimensionless units with the following transformations 
$t \rightarrow \hbar t/\Gamma $, $\vec{x} \rightarrow \hbar \vec{x}/\sqrt{2m\Gamma}$, $\psi  \rightarrow \psi \sqrt{\Gamma/\alpha_1} $ 

\begin{align}
i\frac{\partial\psi_{\pm} (\vec{x})}{\partial t}=& \Big(-\frac{\partial^2}{\partial x^2}-\frac{\partial^2}{\partial y^2}+V(\vec{x})-\Omega_0-\frac{i}{2}+ |\psi_{\pm} (\vec{x})|^2+\alpha_2 |\psi_{\mp} (\vec{x})|^2 +iP_{\pm}(t)\Big) \psi_{\pm}\nonumber\\
&+V_{T}(\vec{x},\phi)\Psi_{\mp}+F_{\pm}
\label{eq:sscheq}
\end{align}
where all the energy scales are normalized by $\Gamma$, $\alpha_2$ by $\alpha_1$, and $F \rightarrow F\sqrt{\alpha_1/\Gamma^3}$.  Initially,  we vary the pump  very slowly to achieve the hysteresis curve given by Fig.~\ref{fig:conti_pot_bi}(e). The characteristic of the hysteresis curves are the same for all the micropillars. Next we fix $F_+$ indicated by the vertical gray line of Fig.~\ref{fig:conti_pot_bi}(e), which corresponds to $F_0=0.22$. Then we apply the $F_-$ pump of the form $F_-=-\eta(2.8\psi_{\text{alive}}+5.2\psi_{\text{dead}})$, where $\eta=0.01$ in the  dimensionless form, $\psi_{\text{alive}}$ corresponds to the steady state solution of Eq.~(\ref{eq:sscheq}) with all the sites alive, and $\psi_{\text{dead}}$ corresponds to the steady state solution with all the sites are dead in presence of $F_+=F_0$. The effect of $F_-$ can be seen clearly in  Fig.~\ref{fig:conti_pot_bi}(f) where the configuration corresponding to  three alive neighbours has the minimum $|\psi_-|^2$. This is equivalent to Fig.~\ref{fig:Fig1}(d). As discussed earlier this is a very crucial step, which ensures that with the choice of proper parameter of $P_\pm$  the states which are initially dead having three
alive neighbours become alive, whereas all other initially dead cells with different neighbours remain dead. Next we vary $P_+$ and $P_-$ simultaneously in presence of $F_+$ and $F_-$ using the periodic boundary conditions. Indeed,  by scanning the space spanned by $P_+$ and $P_-$,  Conway's life can be obtained, which is indicated by the  red dot  in Fig.~\ref{fig:conti_pscan}.  Further in Fig.~\ref{fig:Zoomed_conti_pscan} we have shown that the four configurations lying at the border in Fig.~\ref{fig:conti_pscan} can survive with the 0.02$\%$ fluctuation of $P_\pm$ \cite{Abbaspour2015}.\\


\pagebreak

%

\end{document}